\documentclass[iop,apj]{emulateapj}
\usepackage{epsfig}
\input epsf

\newcommand{\md}{\mbox{$^{\rm o}$}}

\def\gea{\mathrel{\raise.3ex\hbox{$>$}\mkern-14mu \lower0.6ex\hbox{$\sim$}}}
\def\lea{\mathrel{\raise.3ex\hbox{$<$}\mkern-14mu \lower0.6ex\hbox{$\sim$}}}
\newcommand{\Spitzer}{{\bf {\it Spitzer}}}
\newcommand{\ISO}{{\it ISO}}
\newcommand{\Herschel}{{\it Herschel}}

\newcommand{\eg}{e.g.,~}
\newcommand{\ie}{i.e.,~}
\newcommand{\etal}{et al.~}

\newcommand{\um}{\mbox{$\,\mu$m\,}}

\newcommand{\NII}{\mbox{[N\,{\sc ii}]}}
\newcommand{\HII}{H\,{\sc ii}}

\slugcomment{(To appear in the Astrophysical Journal)}

\shorttitle{Mid-infrared dust emissions in M\,81}
\shortauthors{Lu \etal}

\begin{document}

\title{Quantifying the Heating Sources for Mid-infrared Dust Emissions in Galaxies: \\
       The Case of M\,81\footnotemark[$\star$]}

\author{N. Lu\altaffilmark{1}, 
        G. J. Bendo\altaffilmark{2},
        A. Boselli\altaffilmark{3},
	M. Baes\altaffilmark{4},
	H. Wu\altaffilmark{5},
        S. C. Madden\altaffilmark{6}, 
	I. De Looze\altaffilmark{7},
        A. R\'emy-Ruyer\altaffilmark{6, 8},Ê
        M. Boquien\altaffilmark{9},
        C. D. Wilson\altaffilmark{10}, 
        M. Galametz\altaffilmark{11},   
        M. I. Lam\altaffilmark{5}, 
        A. Cooray\altaffilmark{12},    
        L. Spinoglio\altaffilmark{13}, 
        Y. Zhao\altaffilmark{1,14,15}  
       }

\altaffiltext{1}{Infrared Processing and Analysis Center, California Institute of Technology, MS 100-22, Pasadena, CA 91125, USA; lu@ipac.caltech.edu}
\altaffiltext{2}{Jordrell Bank Center for Astrophysics, School of Physics and Astronomy, 
		 University of Manchester, Oxford Road, Manchester M13 9PL, UK}
\altaffiltext{3}{Laboratoire d'Astrophysique de Marseille-LAM, Universit\'e d'Aix-Marseille and CNRS, UMR7326, 
		 38 rue F. Joliot-Curie, F-13388 Marseille Cedex 13, France}
\altaffiltext{4}{Sterrenkundig Observatorium, Universiteit Gent, Krijgslaan 281 S9, B-9000 Gent, Belgium}
\altaffiltext{5}{Key Laboratory of Optical Astronomy, National Astronomical Observatories, Chinese Academy of Sciences, 
		 A20 Datun Road, Beijing 100012, China}
\altaffiltext{6}{Laboratoire AIM, CEA, Universit\'e Paris VII, IRFU/Service d\'Astrophysique, Bat.~709, 91191 Gif-sur-Yvette, France}
\altaffiltext{7}{Sterrenkundig Observatorium, Universiteit Gent, Krijgslaan 281 S9, B-9000 Gent, Belgium}
\altaffiltext{8}{Institut d'Astrophysique Spatiale, CNRS, UMR8617, 91405, Orsay, France}
\altaffiltext{9}{Institute of Astronomy, University of Cambridge, Madingley Road, Cambridge, CB3 0HA, UK}
\altaffiltext{10}{Department of Physics and Astronomy, McMaster University, Hamilton, Ontario L8S 4M1, Canada}
\altaffiltext{11}{European Southern Observatory, Karl-Schwarzschild-Str. 2, D-85748 Garching-bei-MŸnchen, Germany}
\altaffiltext{12}{Department of Physics and Astronomy, University of California, Irvine, CA 92697, USA}
\altaffiltext{13}{Istituto di Astrofisica e Planetologia Spaziali, INAF, Via Fosso del Cavaliere 100, I-00133 Roma, Italy}
\altaffiltext{14}{Purple Mountain Observatory, Chinese Academy of Sciences, Nanjing 210008, China}
\altaffiltext{15}{Key Laboratory of Radio Astronomy, Chinese Academy of Sciences, Nanjing 210008, China}

\footnotetext[$\star$]{{\it Herschel} is an ESA space observatory with science instruments provided by European-led Principal Investigator consortia and with important participation from NASA.}


\begin{abstract}

With the newly available SPIRE images at 250 and 500\um from {\it Herschel Space Observatory},
we study quantitative correlations over a sub-kpc scale among three distinct emission 
components in the interstellar medium of the nearby spiral galaxy M\,81 (NGC\,3031):
(a) $I_{8}$ or $I_{24}$, the surface brightness of the mid-infrared emission observed
in the {\it Spitzer} IRAC 8 or MIPS 24\um\ band, with $I_8$ and $I_{24}$ being 
dominated by the emissions from Polycyclic Aromatic Hydrocarbons (PAHs) and 
very small grains (VSGs) of dust, respectively; (b) $I_{500}$, that of the cold dust 
continuum emission in the {\it Herschel} SPIRE 500\um\ band, dominated by the emission 
from large dust grains heated by evolved stars, and (c) $I_{{\rm H}\alpha}$, 
a nominal surface brightness of the H$\alpha$ line emission, from gas ionized 
by newly formed massive stars.  The results from our correlation study, free 
from any assumption on or modeling of dust emissivity law or dust temperatures, present
solid evidence for significant heating of PAHs and VSGs by evolved stars.  In the case 
of M\,81, about 67\% (48\%) of the 8\um\ (24\um) emission derives its heating from evolved 
stars, with the remainder attributed to radiation heating associated with ionizing stars.

\end{abstract}



\section{Introduction} \label{sec1}

Mid-infrared (5 to $\sim$40\um) emission in the interstellar medium (ISM) of 
normal disk galaxies 
is dominated by (a) broad emission features over 5-15\um, with main
features at 6.2, 7.7, 8.6, 11.3, 12.7\um\ (e.g., Lu \etal 2003), and (b) 
a hot dust continuum longward of $\sim$15\um, which was first detected 
by {\it IRAS} in our own  Galaxy (e.g., Castelaz et al.~1987).
While (a) is widely believed to be the emission bands from Polycyclic 
Aromatic Hydrocarbon molecules (PAHs) of sizes up to $\sim$20\,\AA\ (Puget \& 
L\'eger 1989; Allamandola et al.~1989), (b) is commonly
attributed to the emission from very small grains (VSGs) of dust of 
sizes of $\sim$20 to $\sim$100\,\AA\ (Puget \& L\'eger 1989; D\'esert et al.~1990).
PAHs and VSGs are both heated transiently through absorption 
of single UV/optical photons.  For a given PAH or VSG population and a 
largely fixed shape of heating radiation field,  the spectral shape of 
the PAH or VSG emission should be largely independent
of the intensity of the radiation field (e.g., Draine \& Li 2007).  
As a result, the total flux integrated over their spectrum scales linearly 
with a flux density sampled at any frequency over the spectrum.  A practical
choice is to use the flux densities measured with the {\it Spitzer} IRAC camera
in its 8\um\ band and MIPS camera at 24\um to respectively represent 
the spectrally integrated fluxes of the PAH and VSG emissions. 
We note that the picture described above may not be fully applicable to 
all types of galaxies. For example, for early-type galaxies, the stellar 
emission may be significant at 8\um; for starburst galaxies, the emission
from large dust grains may start to contribute at 24\um\ when their thermal
equilibrium temperature reaches above $\sim$45\,K (e.g., Boulanger et 
al.~1988).

Both the 8 and 24\um\ flux densities have been used widely as tracers of 
current star formation rate (SFR), both in the local universe and at high 
redshifts.  Of these two, the 8\um\ flux is more controversial as to whether 
the heating of the underlying PAHs is primarily derived from 
current star formation:
On one hand, the PAH emission has been shown to correlate with a known
SFR tracer such as the H$\alpha$ line emission, mostly based 
on flux-to-flux correlations (e.g., Roussel \etal 2001; F\"orster Schreiber 
et al.~2004;  Wu \etal 2005; 
Zhu \etal 2008);  on the other hand, it has been well known from {\it COBE} 
(e.g., Dwek \etal 1997),  {\it ISO} (e.g., Lemke \etal 1998; Uchida et al.~1998),
{\it IRTS} (e.g., Chan \etal 2001) and {\it Spitzer} 
(e.g., Lu 2004) that, in our own Galaxy, the PAH emission arises plentifully
in the general ISM and discrete sources devoid of a strong UV radiation 
field.  Additional support for a possible PAH component unrelated to 
current star formation in normal spirals includes striking similarity in 
the large-scale intensity maps of the PAH emission and the cold dust continuum
emission at 850\um\ (Haas \etal 2002), results from various surface brightness
correlation studies based on {\ISO} (e.g., Boselli et al. 2004; Irwin 
\& Madden 2006) and {\it Spitzer} (e.g., Bendo \etal 2008),  as well as 
the fact that either the 8\um\ or the 24\um to the H$\alpha$ flux density 
ratio differs significantly between within or near individual \HII\ 
regions and over galaxies as a whole (e.g., Calzetti et al.~2007; Zhu \etal 2008;
Kennicutt \etal 2009).

A framework that is probably acceptable to both schools of opposite views 
is such that a line-of-sight flux of the PAH emission normally consists of 
two components: a ``warm'' component powered by current star formation and 
a ``cold'' diffuse component powered by the general interstellar radiation 
field dominated by evolved stars.
The controversy described above stems largely from the fact that 
it is practically impossible to unambiguously separate these two components
in a two-dimensional galaxy image as they are not spatially anti-correlated
because the Kennicutt-Schmidt law (Schmidt 1959; Kennicutt 1998) dictates
that the dust/gas distribution should concentrate around star-forming 
regions even if much of the surrounding emission from a particular dust 
population may not be directly related to the current star formation.
One can reduce (but not totally eliminate) this ``degeneracy'' by 
studying dust emission maps at finer linear scales.  

Recent images from {\it Herschel 
Space Observatory} (hereafter {\it Herschel}; Pilbratt \etal 2010) provided 
cold dust emission maps at
linear resolutions as low as $\sim$100 pc in some nearby disk galaxies, 
such as M\,33 (Boquien et al.~2011; Xilouris et al. 2012; Calapa et al.~2014) 
and NGC\,2403 (Jones et al.~2014).  By correlating the 8\um\ emission 
with a sub-millimeter flux density of the cold dust emission or the stellar 
continuum flux density at 3.6\um, these studies generally favor the picture 
that most of the PAH emission in a normal disk galaxy is not directly related 
to the on-going star formation (but also see Croker et al.~2013).  
Most of 
the aforementioned studies inherit one or both of the following limitations: 
they either used a flux density at a fixed wavelength (e.g., 250\um) to 
characterize the cold dust emission or a near-IR stellar continuum (e.g., 
at 3.6\um) to represent the heating radiation field of non-ionizing stars.  
The former could introduce biases as the cold dust emission is always in 
a thermal equilibrium with the heating radiation field and a monochromatic
flux density at a fixed wavelength represents a variable fraction of 
the total (i.e., frequency-integrated) cold dust emission across a galaxy
disk; the latter may bias against the type of non-ionizing stars that are 
most efficient in dust heating when it is used across the entire galaxy.

In this paper, we use an alternative, ``two-component'' analysis to separate
the warm and cold dust emission components across a galaxy disk. This 
method is in principle valid for galaxy images with any angular resolution
and free from either of the potential biases mentioned above.  Our goal is 
to answer 
whether or under what condition the warm PAH component becomes the dominant 
one for a star-forming galaxy as a whole.    Lu \& Helou (2008) pioneered 
this approach using a sample of {\it IRAS} galaxies with available 850\um\ 
fluxes.   Let $F_{\rm warm}$ and $F_{\rm cold}$ be the spectrally integrated
fluxes of the corresponding warm and cold emissions from large dust grains 
in the far infrared (FIR), one can write
\begin{equation}
    F_{\rm PAH} = A\,F_{\rm warm} + B\,F_{\rm cold},
\label{eq1}
\end{equation}
where $A$ and $B$ are two scaling factors to be determined observationally.
$F_{\rm PAH}$ is the spectrally integrated flux of the PAH emission, which
is simply proportional to the flux of any combination of the major PAH 
bands.  Strictly speaking, both A and B, which measure the PAH emission 
relative to the large dust grain emission, are constant under the conditions
of (a) a fixed dust mass spectrum (i.e., a fixed PAH-to-large dust grain
abundance ratio) and (b) a more or less fixed shape of the dust heating radiation
field in each of the cold and warm components (e.g., of similar hardness). 
In reality, both (a) and (b) can certainly 
vary from one galaxy to another and may even vary modestly across a 
galaxy disk.  Nevertheless, it is still meaningful to derive values of A and 
B {\it averaged} over a galaxy disk or galaxies of some similar properties
(e.g. comparable FIR dust color temperatures).

Replacing $F_{\rm PAH}$ with the IRAC 8\um\ flux density and deriving 
$F_{\rm warm}$ and $F_{\rm cold}$ {\it crudely} from the {\it IRAS} 60 and 
100\um and the 850\um fluxes, Lu \& Helou (2008) found statistically that 
the cold PAH component is usually the dominant one except for very actively 
star-forming galaxies [\ie  those with an {\it IRAS} 60-to-100\um\ flux 
density ratio (hereafter referred to as FIR color), $f_{\nu}(60\um) /f_{\nu}(100\um) 
\gtrsim 0.6$].  One implication
from this study is that the fractional PAH emission from the star formation 
component varies from galaxy to galaxy.  If further confirmed, this may 
pose apparent conflict with the constant star-formation fraction of the PAH
emission used in the composite H$\alpha$ and 8\um\ SFR tracer for galaxies
(e.g., Kennicutt et al. 2009).

With the unprecedented sensitivity and improved spatial resolution of 
the SPIRE photometers (Griffin \etal 2010) on board {\it Herschel},
images of nearby spiral galaxies at 250, 350 and 500\um\ with sub-kpc 
linear resolutions have become available.  For normal spiral galaxies, 
these images are usually dominated by the cold dust emission powered 
by evolved stars (e.g., Bendo et al.~2012).  This makes it possible to 
carry out a two-component analysis over the disks of individual galaxies
at a linear resolution limited by the SPIRE images,  not only for 
the PAH emission, but also 
for the VSG dust emission.   In this paper, we present a case study on 
M\,81 (NGC\,3031), one of the nearest normal spiral galaxies.
M\,81 has a morphological type SA(s)ab and an optical disk of $26.9\arcmin 
\times 14.1\arcmin$ (i.e., $R_{25}$ in B; de Vaucouleurs et al.~1991), 
inclined at $59\md$ from face-on (de Blok \etal 2008).
At its distance of 3.63\,Mpc (Karachentsev \etal 2002), the 36.3\arcsec\ 
beam of the full width at half maximum (FWHM) of the SPIRE 500\um\ band
corresponds to 0.64\,kpc. The galaxy is relatively infrared quiescent 
with a FIR color of $\sim$0.26
and has an {\it IRAS} FIR luminosity of $3 \times 10^9\,L\odot$ 
(Rice \etal 1988).   M\,81 is known to host a weak nuclear LINER 
(Heckman 1980), which may have some effect on the nuclear infrared 
emission. However, this should be limited to within one SPIRE 500\um\ beam 
size (Bendo et al.~2012) and have little effect on our statistical 
results that are dominated by the galaxy disk.  In the remainder of 
this paper, we lay out our method in \S2, describe the data used 
in \S3, present our correlation analysis results in \S4 and assess 
potential systematics in \S5. We then discuss some 
astronomical implications in \S6 and finally summarize our main 
conclusions in \S7.

\section{Method} \label{sec2}

We can rewrite eq.~(1) in terms of surface brightnesses.  For the PAH emission, 
we have
\begin{equation}
    I_{8}  = a\,I_{{\rm H}\alpha} + b\,K(T)\,I_{500},
\label{eq2}
\end{equation}
where $I_{8}$ is the surface brightness of the PAH emission in the {\it Spitzer}
IRAC 8\um\ band, $I_{{\rm H}\alpha}$ is the nominal surface brightness of 
the H$\alpha$ line emission, $I_{500}$ is the surface brightness of the dust 
continuum emission in the SPIRE 500\um\ band, and the factor $K(T)$
is defined as such that 
\begin{equation}
   K(T)\,I_{500} = \int^{+\infty}_0 I_{\nu}\,d\nu, 
\label{eq3}
\end{equation}
where the integration is performed in frequency over the cold dust emission 
spectrum of temperature $T$.  While eq.~(3) illustrates the definition of 
$K(T)$ in the context of a single temperature dust emission, {\it $K(T)$ itself 
is a generic K-correction factor that is valid even if the cold dust emission
involves multiple dust temperatures and/or emissivities.}

The first term on the right-hand side of eq.~(2) represents the warm 
component of the PAH emission, under the implicit assumption that 
the H$\alpha$ line emission scales well locally with the far-UV light 
that dominates the heating of the warm dust.
This should be a fairly good assumption for disk galaxies
of normal gas surface brightness and metallicity, such as M\,81, as the dust
and gas are well mixed and the dust absorption of far-UV light is very efficient. 
Observationally, this is backed up by the nearly one-to-one correlation
observed between the PAH and H$\alpha$ surface brightnesses for individual
\HII\ regions in normal galaxies (e.g., Calzetti et al.~2007). (We revisit 
this and other potential systematics in our method in \S5.)

The second term on the right-hand side of eq.~(2) represents the cold
PAH component, under the assumption that $I_{500}$ is dominated by 
the continuum emission of large dust grains heated mainly by evolved
stars.  We show in \S4 that this assumption should be a good one 
in the case of M81, a conclusion that was also reached independently 
by Bendo et al.~(2012).   The factor $K(T)$ in eq. (2)
depends on both cold dust temperature $T$ and dust emissivity $\beta$.
Using the data from and the spectral-energy-distribution (SED) fitting methods 
applied by Bendo et
al. (2010), we see that, at radii greater than 3 kpc in M81, the cold
dust heated by the evolved stars has temperatures of 15-20 K and
$\beta$ values of approximately 2.  For these SED fits, $I_{250}/I_{500}$
increases from $\sim$4 to $\sim$6 with $K(20\,\mbox{K})/K(15\,\mbox{K})
\approx 3$, where $I_{250}$ is the surface brightness in the
SPIRE 250~$\mu$m band.  It will become clear in \S4 that
this empirical approach (as opposed to a full modeling of the quantity
$\int I_{\nu}\,d\nu$) has the advantage that our results are largely free from
any assumption on $T$ or $\beta$. The two constant scaling factors in
eq.~(2), $a$ and $b$, are to be determined from the data.

Similarly we can also write $I_{24}$, the surface brightness of the dust
emission in the {\it Spitzer} 24\um\ band, as follows:
\begin{equation}
    I_{24}  = c\,I_{{\rm H}\alpha} + d\,K(T)\,I_{500},
\label{eq4}
\end{equation}
where $c$ and $d$ are two appropriate scaling factors to be determined.

\section{Data} \label{sec3}

\subsection{{\it Herschel} Images at $250$ and $500$\um} \label{sec3.1}

The SPIRE images used in this paper are the same as those given in 
Bendo \etal (2012) and were from a SPIRE observation (Obs.~ID $= 
1342185538$) under the Very Nearby Galaxy Survey project (PI: C.~D.~Wilson).  
The observation, data reduction and image construction were 
described in detail in Bendo et al.~(2012).  The original images 
have a size of $40\arcmin \times 40\arcmin$ and a flux calibration
uncertainty of $\lesssim$10\% (i.e., {\Herschel} Interactive 
Processing Environment or HIPE, version 5). The overall SPIRE flux calibration
accuracy has since improved. Nevertheless, this has no effect on the correlation 
analysis performed in this paper as the SPIRE detector nonlinearity 
correction remained unchanged.   The adopted spatial resolutions of these 
images are given in Table~1.

Although the original images came with sky subtracted, the surrounding 
regions of M\,81 show patchy cirrus dust emission from our Galaxy 
(Davies \etal 2010). These extended emissions pose some limits on 
the accuracy of sky subtraction.  Nevertheless, a mean residual 
sky background was carefully removed from each SPIRE image.  We 
further smoothed the 250\um\ image to the same spatial resolution 
of the 500\um\ image using a Gaussian kernel with a FWHM of 31.4\arcsec\ 
[$= (36.3\arcsec^2 - 18.2\arcsec^2)^{{1 \over 2}}$].
The smoothed 250\um\ image was then resampled to the same $14\arcsec$
pixel scale as the 500\um\ image.  No color correction was applied 
to either image.

\begin{deluxetable}{lllllll}
\tablenum{1}
\tablecaption{Image Characteristics}
\tablehead{
\colhead{Waveband (\um):}  & \colhead{3.6}   &  \colhead{8}   & \colhead{24}    & \colhead{250}   & \colhead{500} & \colhead{H$\alpha$}}
\startdata
FWHM (arcsec):	           & 1.7$^a$   & 1.9$^a$    & 6.0$^a$   & 18.2$^b$	& 36.3$^b$	& 1.0  \\
Pixel size (arcsec):       & 0.75  & 0.75   & 2.45  & 6	        & 14 	& 2.03 
\enddata
\tablenotetext{a}{See the {\it Spitzer} IRAC or MIPS Instrument Handbook}
\tablenotetext{b}{See the {\it Herschel} SPIRE Handbook}
\end{deluxetable}

\begin{figure}[t]
\centerline{
\psfig{figure=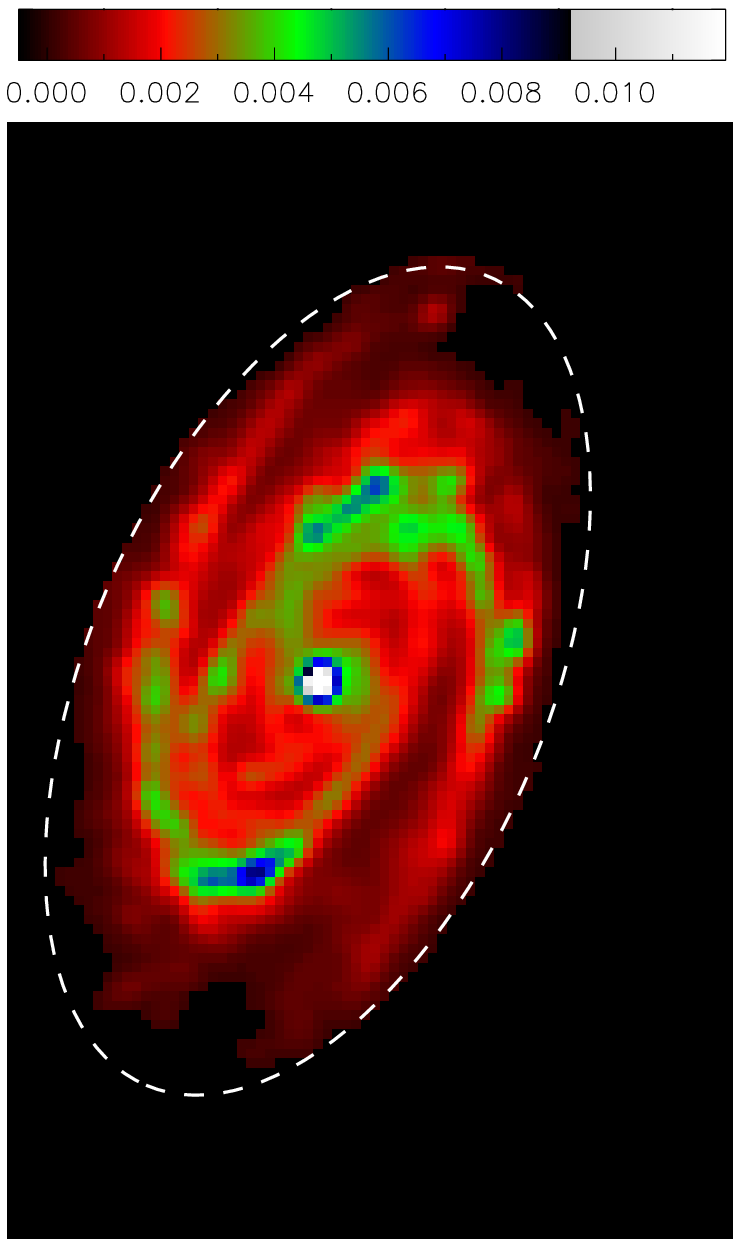,width=4.0cm}
\psfig{figure=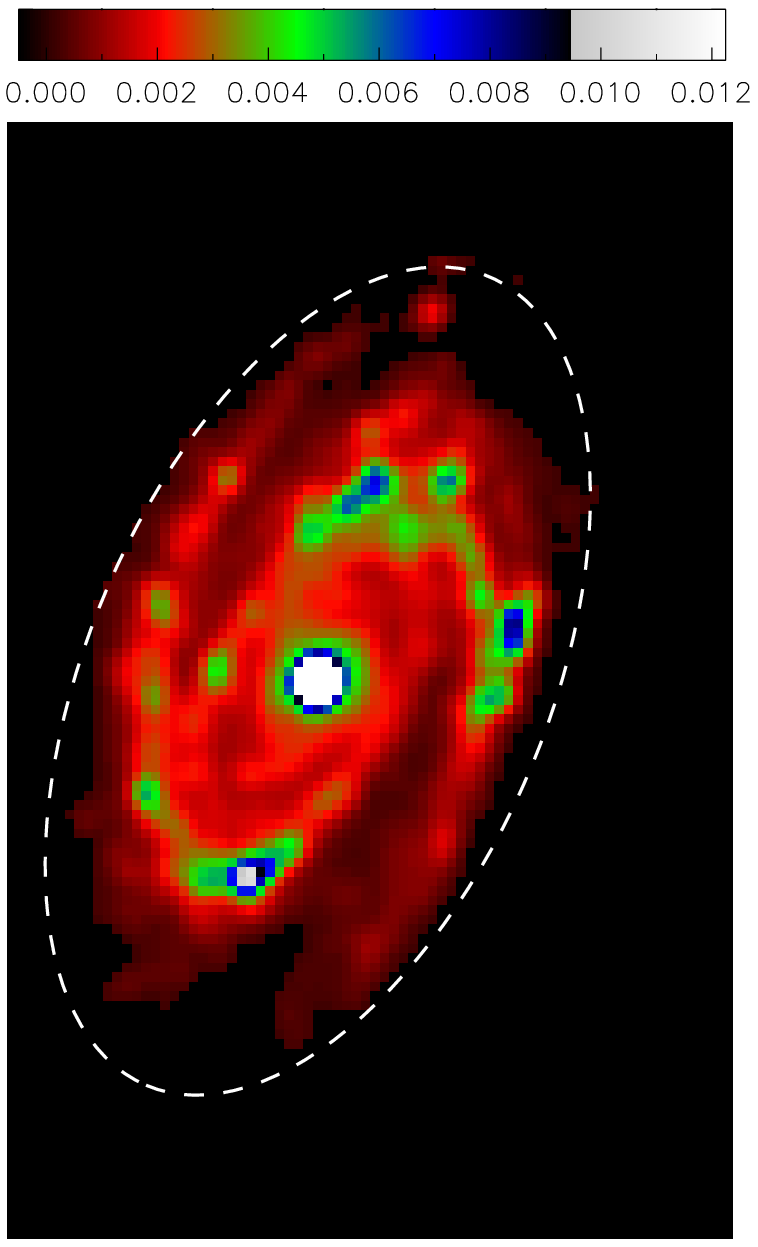,width=4.0cm}
}
\centerline{
\psfig{figure=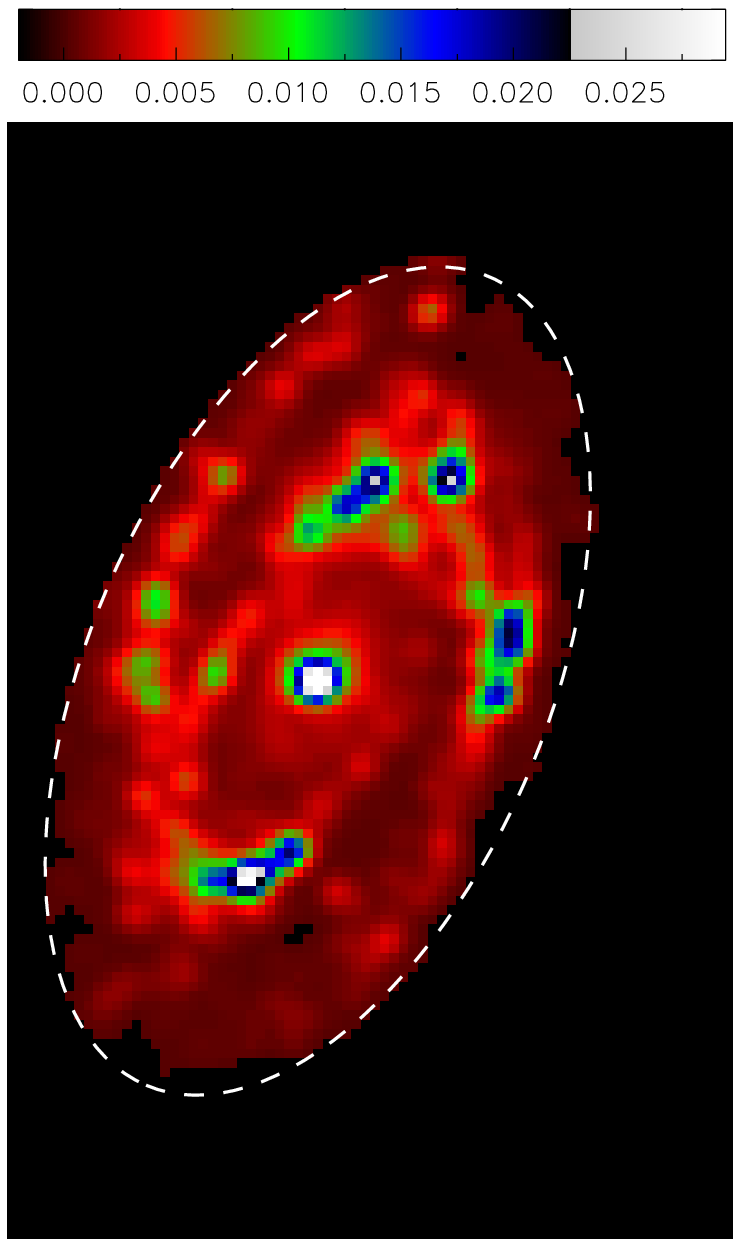,width=4.0cm}
\psfig{figure=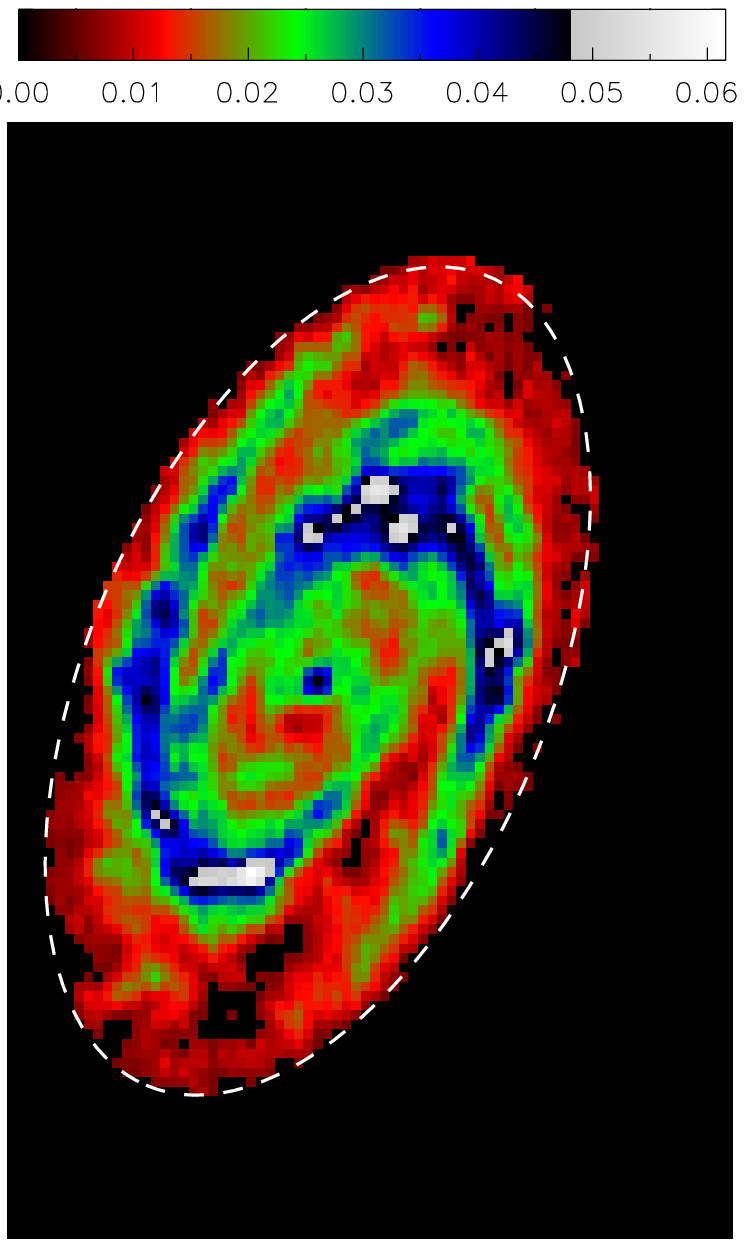,width=4.0cm}
}
\vspace{0.5cm}
\caption{Surface brightness images in linear scale at 8\um\ (top left), 
24\um\ (top right), of the H$\alpha$ line emission (bottom left), and
at 500\um\ (bottom right).
Each image is of 76 pixels (East-West) times 
117 pixels (North-South), with North up and East to the left. 
These images, which have been co-aligned spatially, all have a spatial 
resolution of $\sim$36\arcsec\ and a pixel size of $14\arcsec$. 
The surface brightness is in Jy per pixel (except for the H$\alpha$ image, 
for which it is mJy per pixel) and color-coded according to the color 
bar shown on top of the image. (Note that, in order to show the disk 
structure, the inner galaxy nucleus is saturated in color in the 8\um\, 
24\um\ and H$\alpha$ images.)
Those image pixels with a surface brightness $\le 3\,\sigma$ are deemed
unreliable and shown with a zero intensity here, where $\sigma$ is the sky 
noise in the image.  The ellipse plotted in each image corresponds to a 
face-on circular aperture of 644\arcsec\ in radius. }
\end{figure}

\begin{figure}[t]
\centerline{
\psfig{figure=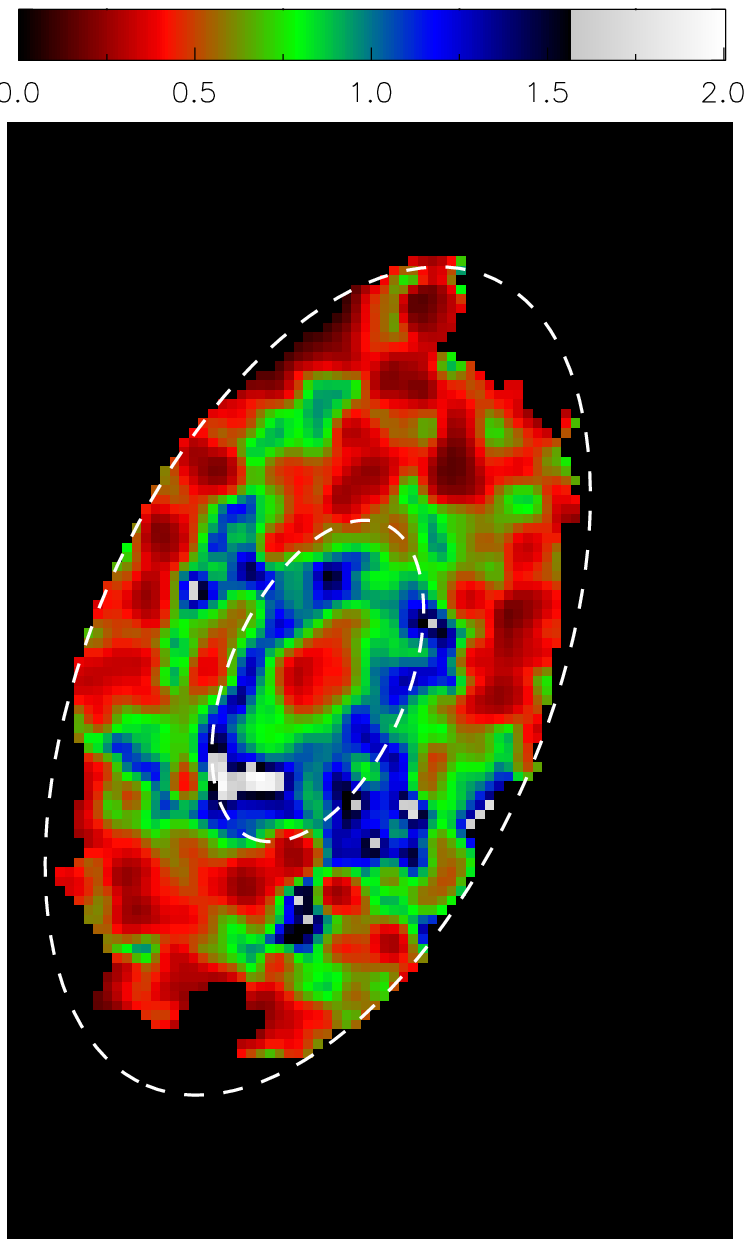,width=4.0cm}
\psfig{figure=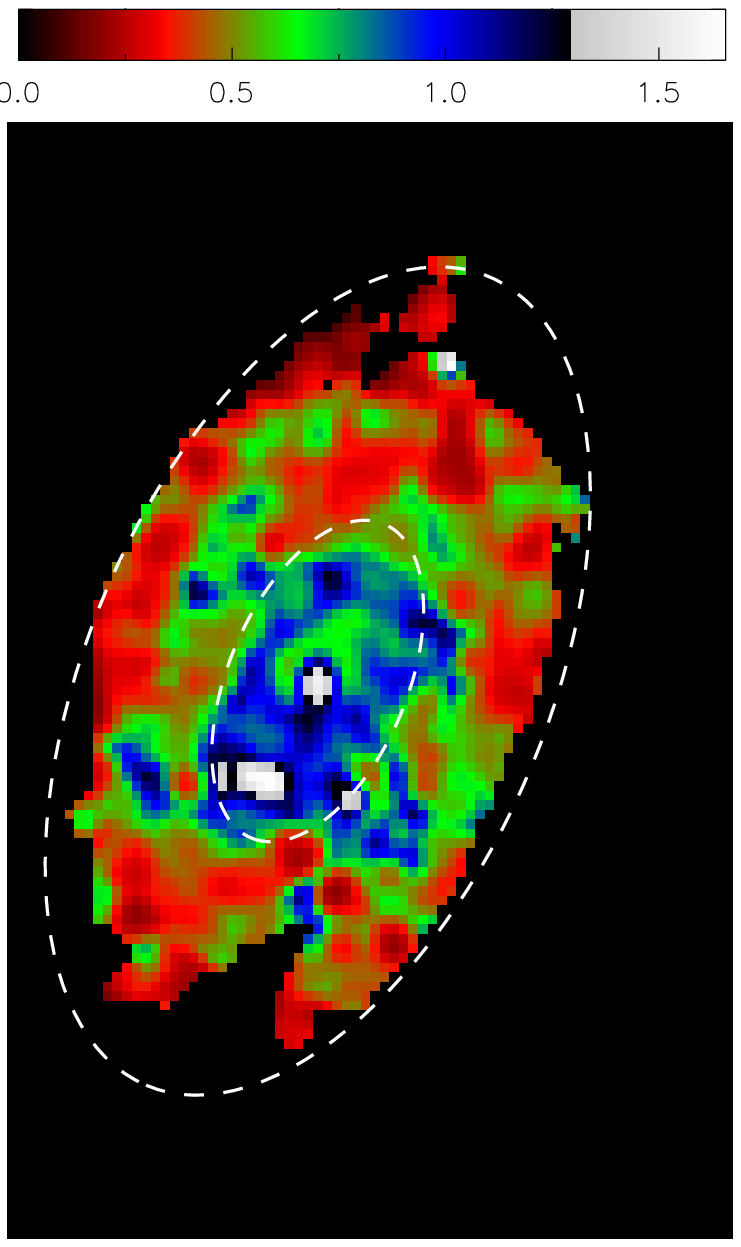,width=4.0cm}
}
\centerline{
\psfig{figure=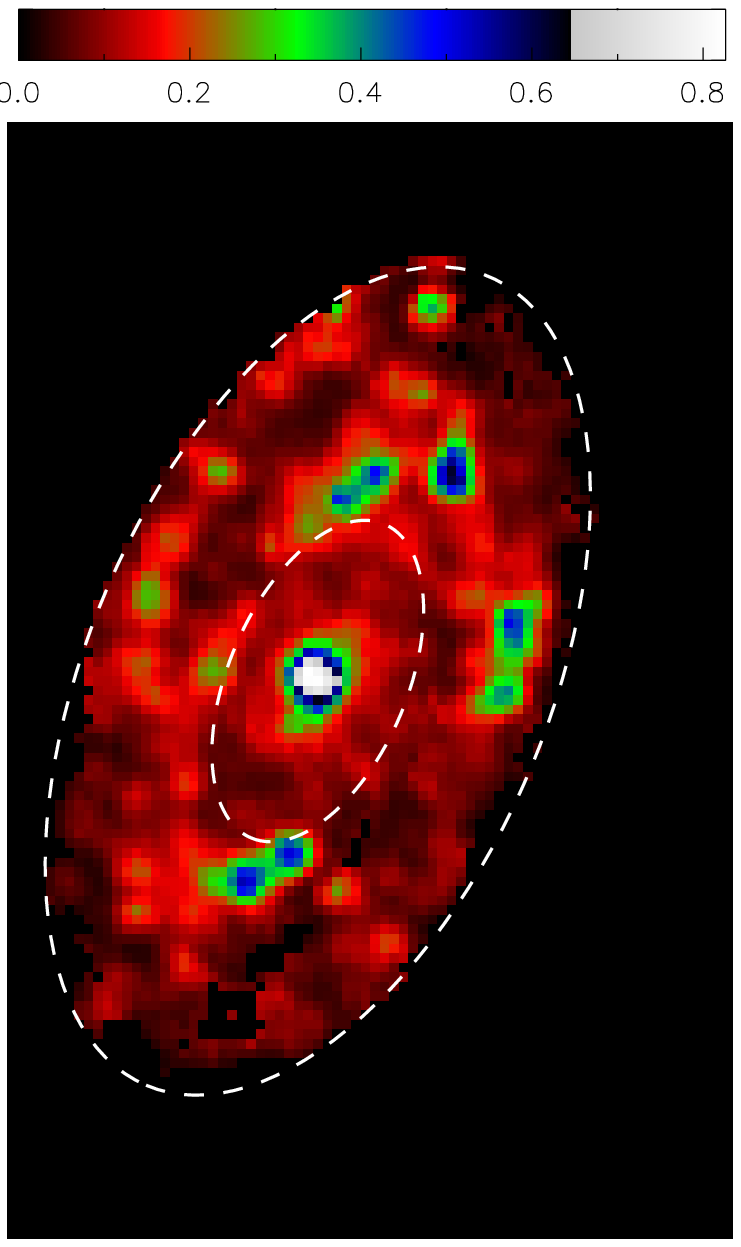,width=4.0cm}
\psfig{figure=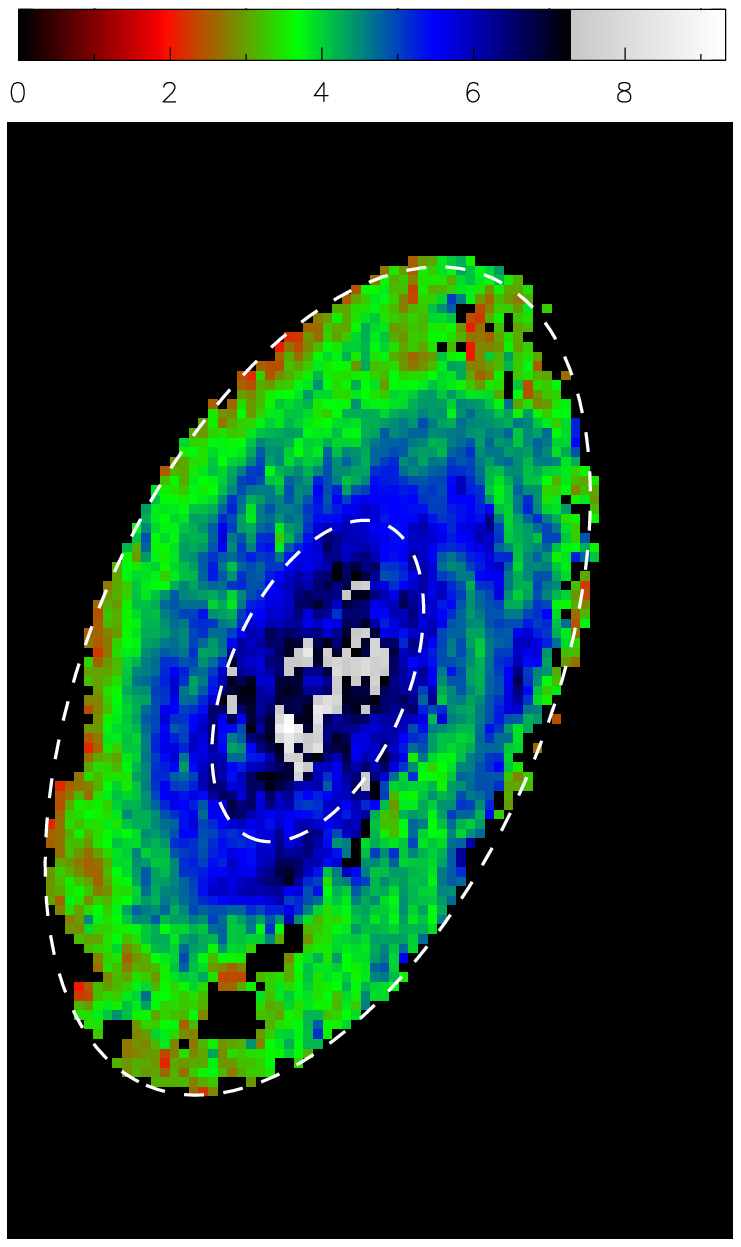,width=4.0cm}
}

\vspace{0.5cm}
\caption{Similar to Fig.~1, but for images of various surface brightness
ratios:  $I_8/(10^3\,I_{{\rm H}\alpha})$ (top left), $I_{24}/(10^3\,I_{{\rm H}\alpha})$ 
(top right), $(10^3\,I_{{\rm H}\alpha})/I_{500}$ (bottom left), and $I_{250}/I_{500}$ 
(bottom right).  The ratios in each image are color-coded according to the color 
bar shown on top of the image.  The two ellipses marked in each image correspond to 
the face-on circles of radii of 250\arcsec\ and 644\arcsec, respectively.  A pixel
of a ratio image is set to zero if either the numerator or denominator 
image has a zero intensity in that image pixel as defined in Fig.~1.
}

\end{figure}

\setcounter{footnote}{15}

\subsection{{\it Spitzer} Images at 8 and 24\um} \label{sec3.2}

The IRAC images at 3.6 and 8.0\um\ and MIPS image at 24\um\ used in 
this paper are the released mosaic images from the {\it Spitzer}
Infrared Nearby Galaxies Survey project (SINGS\footnote{The description
of these images can be found at http://irsa.ipac.caltech.edu/data/SPITZER/SINGS/.}; Kennicutt \etal 2003). 
The images we used here are similar to those in Bendo \etal (2008).
We further multiplied the 3.6 and 8\um\ images by 0.91 and
0.74 (see IRAC Instrument manual, Table 4.8), respectively, to be consistent
with the extended-source flux calibration.

The resulting 3.6\um\ image was smoothed to the spatial resolution of 
the 8\um\ image using a Gaussian kernel with a FWHM of  
$0.85\arcsec$ [$= (1.9\arcsec^2 - 1.7\arcsec^2)^{{1 \over 2}}$] (cf. Table~1).  
The smoothed 3.6\um\ image was used to remove the stellar contribution 
in the 8\um\ image as follows:
\begin{equation}
   I_8 = (I^{\rm tot}_8 -  0.232\,I_{3.6})/0.99, 
\label{eq5}
\end{equation}
where $I^{\rm tot}_8$ and $I_{3.6}$ stand for the observed surface brightnesses
at 8 and 3.6\um, respectively.   The term $0.232\,I_{3.6}$ (see Helou \etal 2004) 
in eq.~(5) accounts for the stellar flux at 8\um\ if all of $I_{3.6}$ is of 
stellar origin.  In reality, the 3.6\um\ flux in disk galaxies is contaminated 
by a dust continuum (\eg Lu \etal 2003; Mentuch et al.~2010).
This contamination is equal to $\sim$5\%\,$I_8$ (Lu 2004).  The denominator
on the right hand side of eq.~(5) corrects for this.
The non-stellar 8\um\ image was subsequently smoothed to the same spatial 
resolution as the SPIRE 500\um\ image, and resampled to a $14\arcsec$ pixel
scale.  This smoothing process used a Gaussian kernel with a FWHM of $36.2\arcsec$ 
[$= (36.3\arcsec^2 - 1.9\arcsec^2)^{{1 \over 2}}$].

As an early-type disk galaxy, M\,81 has a large bulge, which has some stellar
contamination even at 24\um, especially in the low 
surface brightness regions between spiral arms.   This stellar contamination
was removed as follows:  First, the 3.6\um\ image was smoothed to match 
the point spread function (PSF) of the 24\um\ image using an IRAC 3.6 
to MIPS 24\um\ convolution 
kernel\footnote{Available at http://dirty.as.arizona.edu/~kgordon/mips \newline 
/conv$_-$psfs/conv$_-$psfs.html.}
appropriate for a 100\,K blackbody spectrum (see Gordon et al.~2008).  
This kernel function takes into consideration 
the fact that the MIPS 24\um\ PSF deviates from a Gaussian profile.  The kernel 
was first azimuthally averaged before it was convolved with the IRAC 3.6\um\ 
image.   The convolved image was then multiplied by a factor 0.0258 
[$= 0.232\,(8\um/24\um)^2$, with the factor $0.232$ from eq.~(5)] 
and subsequently subtracted from the 24\um\ 
image. The amount of the subtraction accounts for 1-2\% of $I_{24}$ along 
the spiral arms, but could be as large as 10-15\% of $I_{24}$ in 
the low surface brightness regions between the outer spiral arms.
The resulting dust-only 24\um\ image was further smoothed to 
the resolution of the SPIRE 500\um\ band using the same 24-to-500\micron\ 
convolution kernel as in Bendo \etal (2012).   Finally, the resulting 24\um\ 
image was also resampled to a $14\arcsec$ pixel scale.


\subsection{H$\alpha$ Image} \label{sec3.3}

The H$\alpha$ image was taken, via NASA/IPAC Extragalactic Database (NED), 
from Hoopes et al. (2001) with the original image presented 
in Greenawalt \etal (1998).  The image was taken in a filter of 75\AA\ in 
bandwidth, which encompasses the \NII\ doublet.  In M\,81, 
the metallicity varies (mainly radially) slowly across the outer disk
(i.e., over a face-on radius between $\sim$5 and $\sim$11\,kpc), 
but increases more rapidly 
towards the nucleus within the bulge (Garnett \& Shields 1987). We assumed
a constant ratio of \NII/H$\alpha$ for the disk and did not 
subtract \NII\ from the H$\alpha$ image.  We discuss in \S5
the potential systematic effect on our results from possibly higher 
\NII/H$\alpha$ ratios within the bulge.  The image was rotated 
to the correct orientation, smoothed to the resolution of the SPIRE 
500\um\ image using a simple Gaussian kernel based on the FWHM values 
given in Table~1, and resampled to our final pixel size of 14\arcsec.
There are other H$\alpha$ images available (e.g., Boselli \& Gavazzi~2002;
S\'anchez-Gallego et al.~2012).  They are quantitatively very similar to  
the one we used in the paper, at least outside the nucleus.

The original H$\alpha$ image has a flux unit of $2.0 \times 10^{-21}\,$ 
W/m$^2$/arcsec$^2$ (Greenawalt \etal 1998).  We converted the image 
into a nominal surface brightness by dividing each image pixel value
by the H$\alpha$ rest-frame frequency of $4.57 \times 10^{14}\,$Hz. 
No correction for either foreground or internal extinction was done 
in this work.  The foreground and average internal disk extinctions for 
M\,81 are on the order of $A_V \sim 0.2$ and 0.4\,mag, respectively 
(e.g., Perelmuter \& Racine 1995; Schroder et al.~2002) and should have 
no effect on our analysis results.  On the other hand, a significant 
patchiness in the internal extinction could have some systematic 
effect on our analysis.    The inferred internal extinctions towards a 
large number of globular clusters across the entire disk of M\,81 show 
only a moderate variation, within a factor of 2 of the mean value 
(Schroder et al.~2002).  Nevertheless, local internal extinction could 
be significantly higher towards prominent star-forming regions (e.g., 
Kaufman, et al.~1989), resulting in a systematic difference in 
the attenuation of the H$\alpha$ line flux between star-forming and
diffuse regions.  We come back to discuss its effect on our results 
in \S5.

It has been known that star-forming disk galaxies show substantial 
``diffuse'' ionized gas and H$\alpha$ line emission outside bright
\HII\ regions (e.g., Reynolds 1991; Walterbos \& Braun 1994; 
Haffner et al. 2009 and the references therein).  It is generally
believed that this diffuse H$\alpha$ emission is associated with 
ionizing radiation that leaks out of \HII\ regions (e.g., 
Matthis 1986;  Ferguson et al.~1996;  Wood \& Reynolds 1999) although 
some dust scattering of H$\alpha$ photons may also play a role at 
least at high galactic latitudes (e.g., Mattila et al. 2007; Witt
et al. 2010). In this paper,  we follow the notion that the diffuse 
H$\alpha$ line flux originates from current star formation and that 
its associated local UV light powers a warm dust emission component.

All the final images were spatially co-aligned, and have a pixel scale
of 14\arcsec\ and an image size of 76 pixels (East-West) times 117 
pixels (North-South).  The surface brightness units are Jy per pixel 
for all but the H$\alpha$ image, for which the units are mJy per 
pixel.  These images, 
except for the 250\um\ image, are shown in Fig.~1, where only image
pixels with a surface brightness 3 times above the corresponding sky
noise are plotted.  Both {\it Herschel}
250 and 500\um\ images can also be seen in Bendo \etal (2012).

\section{Analysis} \label{sec4}

\subsection{Surface Brightness Ratios} \label{sec4.1}

On the surface, all the four images in Fig.~1 have similar appearances and 
trace each other well qualitatively, except for the nucleus where it is 
relatively faint in the 500\um\ image.  The ellipse shown in each image
corresponds to a face-on angular radius, $R_{\rm face-on} = 644$\arcsec 
($\sim 0.8\,R_{25}$), outside of which $I_{500}$ becomes 
mostly below $3\,\sigma$.   The apparent correlation 
across all the bands over much of the galaxy disk is, to a large extent,
a result of the fact that dust, gas and star distributions are all 
regulated by the spiral pattern.  As a result, one can always find a 
flux-to-flux correlation to a certain degree between any two wavebands.

In Fig.~2 we show surface brightness ratio images of $I_{8}/I_{{\rm H}\alpha}$,
$I_{24}/I_{{\rm H}\alpha}$, $I_{{\rm H}\alpha}/I_{500}$ and $I_{250}/I_{500}$.
The two ellipses shown in each image correspond to the face-on circles of
$R_{\rm face-on} = $ 250\arcsec\ and 644\arcsec (i.e., 4.4 and 11.3\,kpc 
at the galaxy distance), respectively.  
The inner radius was chosen for the following considerations: 
(1) there are no discreet \HII\ regions within this radius (except 
for the nuclear region), and (2) in the outer disk, over $250\arcsec <
R_{\rm face-on} < 644\arcsec$, the bright \HII\ and diffuse regions 
occupy similar ranges in $K(T)$.

While the first three ratio images in Fig.~2 are quite patchy, the image 
of $I_{250}/I_{500}$ is much more smooth, consistent with the claim by 
Bendo \etal (2010, 2012) that the dust emission over these wavelengths 
derives mostly its heating from evolved stars, which are also more smoothly 
distributed than ionizing stars across the galaxy disk.  This is 
further supported by the radial color plots in Fig.~3, where the azimuthal
averages and sample standard deviations in individual radial bins are shown
for the same four surface brightness ratios in Fig.~2.
The averages are normalized by the value in the inner most radial bin;
the sample standard deviations are normalized by the actual (i.e., 
unnormalized) average radio in their corresponding radial bin.
It is clear that the $I_{250}/I_{500}$ ratio is 
much smoother than the other 3 ratios in terms of both radial and azimuthal
variations.  Furthermore, the $I_{250}/I_{500}$ ratio 
shows only marginal change at most across any of those prominent 
disk \HII\ regions visible between the two marked ellipses in 
the image of $I_{{\rm H}\alpha}/I_{500}$.  We found that the reduction 
in $I_{250}/I_{500}$ from within one of these \HII\ regions to 
the general ISM just outside the \HII\ region (but at the same 
galactocentric distance) is no more than 12\%.

Fig.~4 plots the expected color change in $I_{250}/I_{500}$ as a function
of the 250\um\ flux density ratio of a warm dust emission of $T_{\rm warm} 
=$ 30, 40 or 50\,K to a cold dust component of $T_{\rm cold} = 16\,$K.  
Such a cold dust component can reproduce well the observed median ratio
of $I_{250}/I_{500} = 4.5$ over the outer disk when the dust emissivity
scales as $\nu^2$.   Two-component SED fits to 
the FIR dust emission in normal galaxies generally result in $T_{\rm warm}$ 
between 30 and 40\,K (e.g., Bendo \etal 2003), which correspond to 
the region between the solid and dashed curves in Fig.~4. These curves show 
that a Y-axis value of 1.14 [$= 1/(1-12\%)$; i.e., a 14\% \HII-related
enhancement relative to the diffuse region just outside the \HII\ region] 
corresponds to an X-axis reading
of less than 0.3.  As a result, the warm dust emission associated 
with the current star formation in M\,81 contributes to less than 
$\sim$23\% of the total observed flux at 250\um based on Fig.~4.  
At 500\um, this contribution should be less than $\sim$13\%.  
In the remainder of this paper, we regard $I_{500}$ as being only 
associated with large dust grains heated by evolved stars.

\begin{figure}
\centerline{
\psfig{file=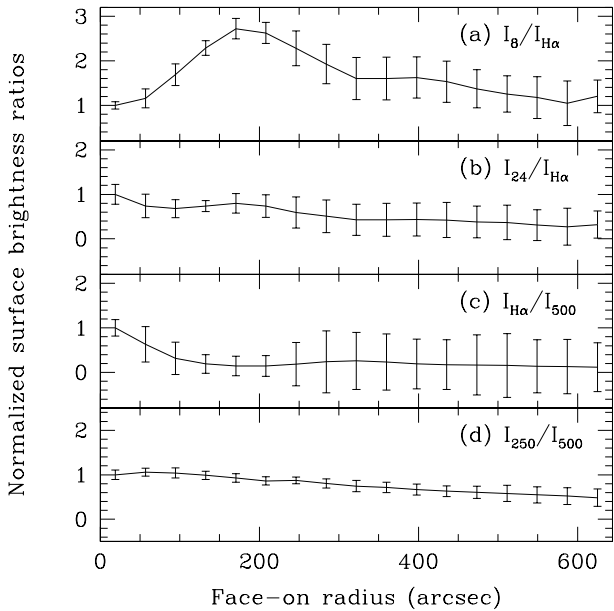, width=9.0cm, angle=0, bbllx=18, bblly=144, bburx = 550, bbury=690}
}
\vspace{-0.2in}
\caption[]{
	Radial plots of the azimuthally averaged surface brightness ratios: 
        (a) $I_8/I_{{\rm H}\alpha}$,
        (b) $I_{24}/I_{{\rm H}\alpha}$, (c) $I_{{\rm H}\alpha}/I_{500}$ and (d)
        $I_{250}/I_{500}$.  Each radial profile has been normalized by its average 
        in the inner most radial bin. The error bars shown are the sample standard 
        deviation normalized by the actual observed average in the corresponding 
	radial bin. The plots have the same 
        vertical range so that the relative ratio variations can be visually 
        compared among these plots.
}
\end{figure}

\begin{figure}
\centerline{
\psfig{file=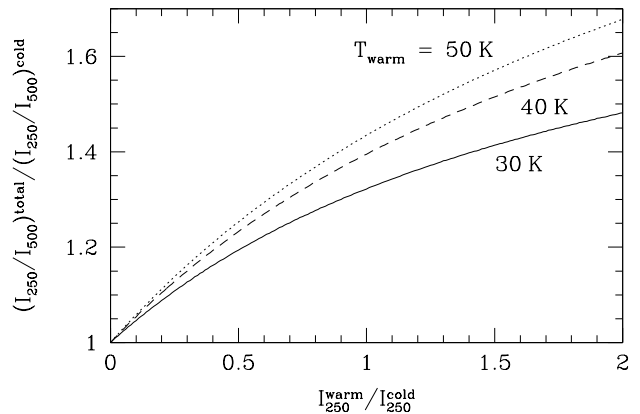, width=9.0cm, angle=0, bbllx=18, bblly=144, bburx = 550, bbury=718}
}
\vspace{-1.2in}
\caption[]{
	Plot of the enhancement to $I_{250}/I_{500}$ from a warm dust emission of temperature 
	$T_{\rm warm}$ over a cold dust emission of $T_{\rm cold} = 16\,$K, as a function of 
	the warm-to-cold surface brightness ratio at 250\um.  We assumed a common dust
      emissivity that scales with $\nu^2$.  As labeled, the plotted curves used 
      $T_{\rm warm} = 30$, $40$ or $50\,$K.
}
\end{figure}

Fig.~2 also reveals an apparent anti-correlation between either $I_{8}/I_{{\rm H}\alpha}$
or $I_{24}/I_{{\rm H}\alpha}$ and $I_{{\rm H}\alpha}/I_{500}$, in the sense that 
local peaks in the last ratio correspond to local dips in either of the former
ratios, suggesting that there are factors other than $I_{{\rm H}\alpha}$ in powering
$I_{8}$ or $I_{24}$.
In particular,  both the $I_{8}/I_{{\rm H}\alpha}$ and $I_{24}/I_{{\rm H}\alpha}$ 
images have peaks in the bulge/inner disk region, between the outer disk, where 
the major spiral arms and prominent 
\HII\ regions reside, and the circumnuclear region, where an enhanced H$\alpha$ 
emission is seen. 
This bulge/inner disk area is devoid of discrete H$\alpha$ emission of high surface 
brightness, yet, the original high-resolution IRAC 8\um\ image shows clearly PAH 
emission in spiral patterns (see Willner \etal 2004).   Taking at face value, 
the ratio images in Fig.~2 suggest that evolved stars play a significant role in 
heating both PAHs and VSGs in the case of M\,81.

\subsection{Correlation Analysis} \label{sec4.2}

We can analyze Fig.~2 in a more quantitative way via eqs.~(2) to (4).  To do so, 
we first resampled all the images to a pixel size of 36\arcsec, roughly equal to 
the spatial resolution of the SPIRE 500\um\ band. This is to ensure that individual 
image pixels are statistically independent from each other.  With all the images
having the same flux units of Jy per pixel, Fig.~5 is a simple pixel plot of 
$I_{8}/I_{500}$ {\it vs.} $I_{{\rm H}\alpha}/I_{500}$ using all those 36\arcsec\ 
image pixels between $R_{\rm face-on}$ of 250\arcsec\ and 644\arcsec, for which
the 3 surface brightness values involved are all above 3 times their corresponding 
background noise.  This outer disk hosts prominent \HII\ regions, along 
the major spiral arms, which should provide a significant or even dominant heating 
source for the mid-IR dust emissions.  Another practical consideration of using 
only the outer disk in Fig.~5 is to maintain more or less the same (and finite) range 
in the values of $K(T)$ between the diffuse regions and the \HII-dominated regions, 
in order to obtain an unbiased slope from fitting eq.~(2) or (4) to the data.

\begin{figure}[t]
\centerline{
\psfig{file=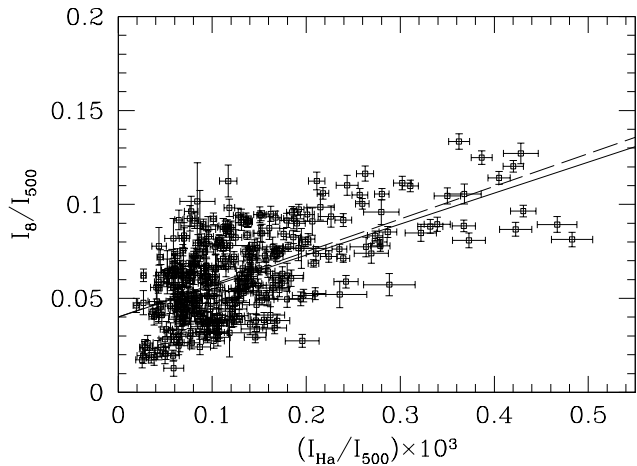, width=9.5cm, angle=0, bbllx=18, bblly=144, bburx = 550, bbury=718}
}
\vspace{-3cm}
\caption[]{
	Plot of $I_{8}/I_{500}$ {\it vs}. $I_{{\rm H}\alpha}/I_{500}$ for all the 36\arcsec\ pixels 
	in the outer disk between the face-on radii of 250\arcsec and 644\arcsec, for which all 
        the involved surface brightness values are greater than 3 times their respective 
        image background noise.  The error bars shown are at $1\sigma$.  The solid 
        and dotted lines are respectively least-squares fits from a simple regression minimizing 
        the residuals along the vertical axis and a full regression that minimizes the residuals 
        along both axes.
       	}
\end{figure}

Indeed, the outer-disk data points in Fig.~5 span a range of a factor of 
$\gtrsim$10 in terms of $I_{{\rm H}\alpha}/I_{500}$. In contrast, $I_{250}/I_{500}$ 
remains smooth and varies only by a factor of $\sim$2 (see Fig.~2).  
We therefore applied least-squares fits of eq.~(2)
to the data points in Fig.~5, resulting in the two lines shown.  
The solid line is from a simple regression by
minimizing the vertical residuals, while the dotted line is from a full
regression by minimizing the residuals along both axes. Within the uncertainties, 
both the fits agree with each other and can be represented by 
the following result:
\begin{equation}
    I_{8}  = (0.17 \pm 0.01)\times 10^3\,I_{{\rm H}\alpha} + (0.040 \pm 0.002)\,I_{500}.
\label{eq6}
\end{equation}

\begin{figure}[t]
\centerline{
\psfig{file=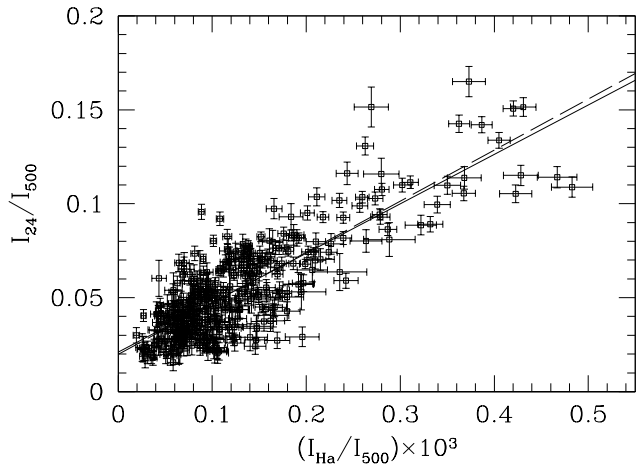, width=9.5cm, angle=0, bbllx=18, bblly=144, bburx = 550, bbury=718}
}
\vspace{-3cm}
\caption[]{Same plot as in Fig.~5, but of $I_{24}/I_{500}$ {\it vs}. $I_{{\rm H}\alpha}/I_{500}$.
       	}
\end{figure}

We also carried out the same analysis on the 24\um\ emission in Fig.~6 via eq.~(4). 
Fig.~6 is similar to Fig.~5, but using $I_{24}/I_{500}$ instead.  
The least-square fits shown in the figure can be represented by
\begin{equation}
    I_{24}  = (0.27 \pm 0.01)\times 10^3\,I_{{\rm H}\alpha} + (0.021 \pm 0.002)\,I_{500},
\label{eq7}
\end{equation}

\begin{deluxetable}{lccc}
\tablenum{2}
\tablewidth{0cm}
\tablecaption{Fractional flux contributions from the Warm Dust Component}
\tablehead{
\colhead{Region} & \colhead{$I_{{\rm H}\alpha}/I_{500}$}  & \colhead{$I_8{\rm (warm)}/I_8{\rm (tot)}$} & \colhead{$I_{24}{\rm (warm)}/I_{24}{\rm (tot)}$}}
\startdata
Diffuse      & $0.04 \times 10^{-3}$             & $0.15 (\pm 0.01^a)$                  & $0.34 (\pm 0.02^a)$ \\
\ion{H}{2}   & $0.40 \times 10^{-3}$             & $0.63 (\pm 0.02^a)$                  & $0.84 (\pm 0.01^a)$
\enddata
\tablenotetext{a}{The errors were estimated using the parameter uncertainties in eq.~(6) or (7)}
\end{deluxetable}

It is interesting to see, via eq.~(6) or (7), how the warm and cold dust 
components compare with each other at different locations in the outer 
disk of the galaxy.   Table~2 compares the fractional flux contributions
at 8 or 24 \micron\ from the warm dust component between two locations 
that differ by a factor of 10 in terms of $I_{{\rm H}\alpha}/I_{500}$.
As evident in Fig.~5 or 6, these two locations represent respectively 
a diffuse region with the lowest and a bright \HII\ region
with nearly the highest values of $I_{{\rm H}\alpha}/I_{500}$ observed.
For the 8\um\ emission, the fractional 
flux contribution from the warm component varies from 15\% in such a 
diffuse region to 63\% in such a bright \HII\ region; 
for the 24\um emission, this variation is 34 to 84\%.   This shows 
clearly that the current star formation dominates {\it both} PAH and 
24\um\ dust emissions in or around those bright \HII\ regions in 
M\,81.  In contrast, in the diffuse region considered in Table~2, this 
star formation component only accounts for $\sim$15\% and 34\% of 
the total flux at 8 and 24\um, respectively.

\subsection{Variations in $K(T)$} \label{sec4.1}

The scatter in Fig.~5 or 6 is much larger than what the statistical 
errors (shown in the figure) can account for. Fig.~7 demonstrates 
that much of this scatter is due to the variation of the factor $K(T)$, 
defined in eq.~(3), by showing the image pixels in plots of 
the normalized $I_{8}/I_{500}$ or $I_{24}/I_{500}$ as a function
of $I_{250}/I_{500}$, where the normalization was done by dividing
the observed ratio by the mean value given in eq.~(6) or (7). 
For example, $(I_8/I_{500})_{\rm normalized} = (I_8/I_{500}) /
(0.17\times10^3\,I_{{\rm H}\alpha}/I_{500} + 0.040)$. Therefore,
an image pixel lying above (or below) the mean fit in 
Fig.~5 would have the normalized ratio $(I_{8}/I_{500})_{\rm normalized}
> 1$ (or $<1$).  In Fig.~7, we showed the image pixels from both 
the outer disk (i.e., the black squares in the figure) and inner 
disk (red squares) delineated at $R_{\rm face-on} = 250\arcsec$.
Again, only those image pixels with the appropriate surface brightnesses 
3 times greater than their respective uncertainties are plotted here.  
For the outer disk data points, as 
$I_{250}/I_{500}$ [or $K(T)$)] increases, the normalized $I_{8}/I_{500}$
or $I_{24}/I_{500}$ ratio increases from being $< 1$ to $>1$.  This 
trend contributes to much of the scatter seen in Fig.~5 or 6. 
If we had included the inner disk image pixels in Fig.~5 or 6, these 
data points would be farther above our mean fit to the outer disk 
data points as the inner data points have even larger values of $K(T)$. 
In principle, one can divide
a galaxy disk into successive radial annuli and fit eq.~(2) or (4) to 
individual annuli independently as long as $I_{{\rm H}\alpha}/I_{500}$
in each annulus covers enough dynamic range to make such a fit feasible.
The fit from each radial annulus, of a different average value of $K(T)$, 
should give more or less the same slope if 
the scaling factor ``a'' in eq.~(2) [or ``c''
in eq.~(4)] does not depend on the galactocentric radius. In the case
of M\,81, it is not feasible to make a separate two-component analysis 
in the inner disk which contains far fewer image pixels that are also
distributed over a much larger range in $K(T)$ (as evidenced by a larger 
vertical spread for the inner disk data points in Fig.~7).

\begin{figure}[t]
\centerline{
\psfig{file=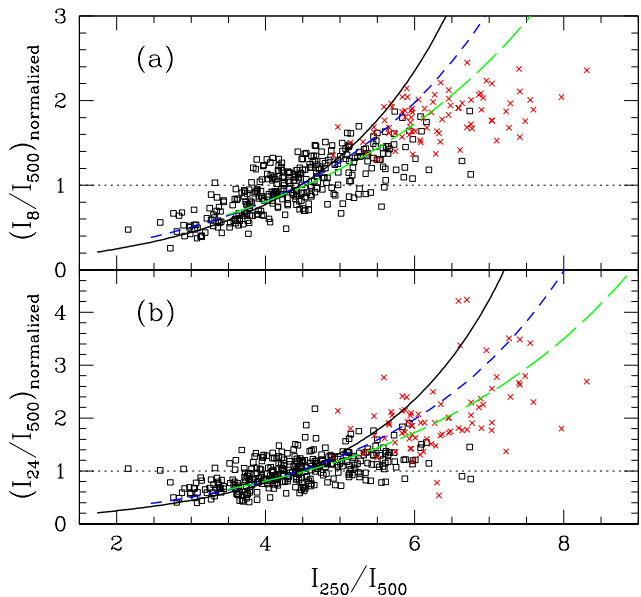, width=9.5cm, angle=0, bbllx=48, bblly=144, bburx = 540, bbury=718}
}
\vspace{-2.0cm}
\caption[]{Plots of the image pixels of (a) $I_{8}/I_{500}$ and (b) $I_{24}/I_{500}$, 
           each normalized by their corresponding mean value given by eq.~(6) or (7), 
           as a function of $I_{250}/I_{500}$.  Those from the outer and inner parts of
	   the galaxy disk, divided at the face-on circle of radius $R_{\rm face-on} 
           = 250\arcsec$, are shown by squares and crosses (colored red in the online
	   version), respectively.  Only those pixels with all the involved surface 
	   brightnesses 3 times greater than their respective uncertainties are plotted.
           The solid, short-dashed 
           and long-dashed curves (colored in black, blue and green respectively 
	   in the online version)
           are predictions from a Planck function modified by an 
           emissivity that scales with frequency $\nu$ as $\nu^{\beta}$ for $\beta =$ 
           2, 2.5 and 3, respectively.  The left end of each model curve corresponds 
	   to the lower limit of 10\,K in dust temperature.  All the curves were 
           normalized to unity at $I_{250}/I_{500} = 4.5$. The horizontal dotted line
	   marks where the mean fit of eq.~(6) or (7) is.
          }
\end{figure}

The three curves in each plot of Fig.~7 correspond to the predicted $K(T)$ values from 
a Planck function 
modified with a dust emissivity $\propto \nu^{\beta}$ with $\beta = 2$, 2.5 and 3, 
respectively. These values were simply chosen for illustration purpose although 
the $\beta = 2$ case was favored in the SED analysis of Bendo \etal (2010).  The left 
end of each curve corresponds to a dust temperature of 10\,K.  For the $\beta = 2$
model, the dust temperatures are 12.6, 14.7, 17.0, 19.5, 22.5 and 26.0\,K at 
$I_{250}/I_{500} = 3$, 4, 5, 6, 7 and 8, respectively.  The similar temperature sets
for the $\beta = 2.5$ and 3 models are (10.8, 12.3, 13.7, 15.2, 16.9, 18.6\,K) 
and (N/A, 10.5, 11.6, 12.6, 13.7, 14.7\,K), respectively.   All the curves are 
normalized to unity at $I_{250}/I_{500} = 4.5$.  

While none of these curves can fit the data points over the entire galaxy, they 
do follow the trend of the data points in the outer disk, consistent with our 
claim that a large part of the scatter seen in Fig.~5 or 6 can be accounted for 
by the variation in $K(T)$.  The inner disk data points (i.e., the crosses
in Fig.~7) tend to fall below the model curves.  This is not necessarily suggestive 
of a greater value of $\beta$.  As further discussed in \S5, this could also be 
a result of a greater contamination of the \NII\ emission in our H$\alpha$ flux 
within the inner disk, where a significantly higher metallicity was observed.

\begin{deluxetable}{llll}
\tablenum{3}
\tablewidth{0cm}
\tablecaption{Flux Densities of Cold and Warm Dust Components}
\tablehead{
\colhead{Image\ \ \ }  &  \colhead{Total\ \ \ \ }    & \colhead{Warm\ \ \ \ \ \ \ \ }         & \colhead{Cold\ \ \ \ \ \ \ \ \ } \\
\colhead{}             &  \colhead{(Jy)\ \ \ \ }             & \colhead{(Jy)\ \ \ \ \ \ \ \ }                 &  \colhead{(Jy)\ \ \ \ \ \ \ \ \ } \\
\colhead{(1)\ \ \ }    &  \colhead{(2)\ \ \ \  }              & \colhead{(3)\ \ \ \ \ \ \ \ }                  &  \colhead{(4)\ \ \ \ \ \ \ \ \ }} 
\startdata
Outer disk$^a$: \\
\ \ 8\um             & 3.58\ \ \ \        & 1.31 $(\pm 0.08)$ \ \ \ \       & 2.20 $(\pm 0.11)$\ \ \ \  \\
\ \ 24\um            & 3.22\ \ \ \        & 2.08 $(\pm 0.08)$\ \ \ \        & 1.15 $(\pm 0.11)$ \ \ \ \  \\
\ \ H$\alpha$        & 0.0077$^b$\ \ \    & 0.0077$^b$\ \ \    &0.0 \ \ \ \  \\
\ \ 500\um           & 54.96\ \ \ \       & 0.0 \ \ \ \        &54.96\ \ \ \  \\
\\
Total galaxy$^c$: \\
\ \ 8\um             & 5.03\ \ \ \        & 1.65\ \ \ \       &3.38\ \ \ \  \\
\ \ 24\um            & 5.01\ \ \ \        & 2.62\ \ \ \       &2.39\ \ \ \  \\
\ \ H$\alpha$        & 0.0097$^b$\ \ \    & 0.0097$^b$\ \ \   &0.0 \ \ \ \  \\
\ \ 500\um           & 67.07\ \ \ \       & 0.0 \ \ \ \       & 67.07\ \ \ \  
\enddata
\tablenotetext{a}{Defined to be the disk between the face-on radii of 250\arcsec\ and 644\arcsec.}
\tablenotetext{b}{The surface brightness of the H$\alpha$ image is defined to be
the line flux divided by the rest-frame line frequency of $4.57 \times 10^{14}\,$Hz.}
\tablenotetext{c}{Defined to be the galaxy within the face-on radius of 644\arcsec.}
\end{deluxetable}


\subsection{Total Fluxes} \label{sec4.4}

We derived a total galaxy flux in each image by summing over all image pixels
within $R_{\rm face-on} = 644\arcsec$, including those with a surface brightness
less than 3 times the sky noise.   Similarly, an integrated flux for the outer
disk of $250\arcsec < R_{\rm face-on} \le 644\arcsec$ was also derived.  These fluxes
are shown in Column~(2) of Table~3.    For the outer disk, eq.~(6) or (7) was used
to derive the 8 or 24\um\ fluxes for the warm and cold components, given in Columns~(3)
and (4), respectively. The associated errors were based on the parameter uncertainties
in eq.~(6) or (7). Since eq.~(6) or (7) was fit to this outer disk, it is not 
surprising that the sum of the warm and cold component fluxes in the table agrees 
with the measured total flux in Column~(2) for either the 8 or 24\um\ emission.
These results show that, in the outer disk of M\,81, $63\% (\pm 2\%)$ of the PAH 
emission is not directly associated with the current star formation.  For the 24\um\ 
emission, this fraction is $35\% (\pm 2\%)$.

To derive the flux of the cold or warm dust component for the total galaxy, we 
also need to calculate the contribution from the inner disk/bulge region with
$R_{\rm face-on} < 250\arcsec$. To this end, we assumed that the scaling constant 
{\it for the warm component} in eq.~(6) or (7) remains valid in the inner 
disk/bulge region, and estimated the contribution from the current star formation
to $I_8$ or $I_{24}$ using the observed H$\alpha$ flux. Because of this additional
assumption, the results for the inner disk region may be subject to additional 
systematics in the H$\alpha$ flux as discussed in \S5. We therefore listed 
in Table~3 the results for the outer disk separately from those for the entire 
galaxy. In the later case,  $\sim$67\% of the total PAH emission in M\,81 is not
directly associated with 
the current star formation.  For the 24\um\ emission, this fraction is $\sim$48\%.
Clearly, the inclusion of the inner disk and the bulge, where the surface density 
of evolved stars is high, reduces the overall contribution from the current star 
formation.


\section{Assessment of Systematics} \label{sec5}

As we mentioned in \S1, our analysis improved over many of the previous studies
on the same subject in two areas: (a) we used a spectrally integrated flux of 
the cold dust emission instead of a monochromatic flux density; (b) we made 
no assumption on whether the 3.6\um\ flux density is a good proxy for the dust 
heating from evolved stars\footnote{The 3.6\um\ flux density was used here to 
{\it only} estimate the stellar continuum at 8\um.  At wavelengths longer than 
$\sim$3.6\um, the spectral shape of the stellar continuum of a normal galaxy 
depends little on star formation history or metallicity (Helou 
\etal 2004).  From 8 to 24\um, our simple Rayleigh-Jeans approximation of 
the stellar continuum should be sufficient.}.
Furthermore, our analysis is empirical, without prior assumption for the dust 
emissivity law or dust temperatures.  This differs from most conventional SED 
analyses in the literature.  On the other hand, 
our analysis anchored on two basic requirements and may be subject to some 
systematics.  We discuss these two issues separately below.

As introduced in \S1, our first basic requirement is that $I_{500}$ is dominated
by the emission from large dust grains heated by evolved stars.  This was shown
to be the case in \S4.1 in this paper and in Bendo \etal (2012). 
Furthermore, we also require that the (extinction-corrected) H$\alpha$ flux 
scales linearly with the far-UV photon flux that dominates the heating of 
warm dust.  This requirement translates to (a) a well mixed gas and dust over 
our spatial resolution element ($\sim$36\arcsec\ or 0.64\,kpc) and (b) a fixed 
initial mass function (IMF) of the current star formation.  While this remains 
as an assumption
in principle in this paper, it is a reasonable one based on the observations 
that the PAH and H$\alpha$ fluxes scale almost linearly with each other for 
individual \HII\ regions in nearby normal galaxies (e.g., Calzetti \etal 2007). 
Nevertheless, even if the IMF varies from one star-forming region to another, 
our results should remain statistically valid for the outer disk of M\,81. 
This also holds true if the PAH-to-large dust grain abundance ratio varies across
the galaxy disk.

In contrast, our analysis results may be subject to a number of potential systematics 
in the H$\alpha$ flux as we already alluded to in \S3.3.  In the outer disk, 
the internal optical extinction follows gas/dust distribution and generally peaks 
around bright \HII\ regions, where the largest surface brightness of the H$\alpha$
emission is observed.  As a result, if we had applied some internal extinction 
correction to the H$\alpha$ fluxes, the fit of eq.~(6) or (7) to the data would 
be somewhat flatter, making its Y-axis intercept even greater in Fig.~5 or 6.  
In other words, any correction for the internal extinction on $I_{{\rm H}\alpha}$ 
would likely result in even a greater cold component in $I_8$ or $I_{24}$.

Another possible systematic issue is the elevated metallicity in the galaxy bulge. 
If we had taken into account the fact that the \NII\ line might be relatively 
brighter within the galaxy bulge, the warm component fluxes for the inner disk,
thus for the {\it total} galaxy in Column (3) of Table~3, would have been somewhat 
overestimated.  As a result, the counterpart fluxes of the cold component in 
Column~(4) would have been underestimated accordingly.

Furthermore, a potential controversy exists regarding whether the circumnuclear 
H$\alpha$ emission in M\,81 is derived from current star formation (see Devereux
et al.~1995).  If it is not related to the ongoing star formation, the inner disk 
contribution to the warm dust components in Column~(3) in Table~3 should be 
further reduced, making the overall cold dust component even more significant. 

In summary, all these potential systematics in the H$\alpha$ image can only imply 
that our estimates of the cold-component fractional fluxes at both 8 and 24\um\ 
in Table~3, especially in the inner disk/bulge region, might be somewhat 
underestimated.

\section{Discussion} \label{sec6}

Our correlation analysis showed that the 24\um\ emission from VSGs is 
indeed a better SFR tracer than the 8\um\ PAH emission as the former
is 60\% more sensitive to ionizing stars and 50\% less sensitive 
to evolved stars than the latter, based on the slopes in eqs.~(6) and
(7).  In fact, Table~3 shows that the 24\um\ flux in the outer disk 
of M\,81 is indeed modestly dominated by the current star formation.
This is in agreement with the latest consensus in the literature.
On the other hand, $I_{24}$ still derives nearly half of its emission 
from the general ISM in the case of M\,81 when integrated over 
the galaxy surface, suggesting that even the 24\um\ emission could arise 
significantly from VSGs heated by evolved stars for a galaxy as whole. 
The relatively large stellar mass of M\,81 may play a role, as
implied by studies of other early-type disk galaxies (e.g., Sauvage \& 
Thuan 1992; Engelbracht et al. 2010).   Tabatabaei \& Berkhuijsen
(2010) reached a similar conclusion on the heating source for the 24 
\micron\ emission in M\,31, another nearby IR-quiescent disk galaxy.

It is interesting to compare our M\,81 results with the statistical 
results from analyzing a sample of {\it IRAS} galaxies in
Lu \& Helou~(2008).  Their statistical conclusion is that, for 
galaxies with quiescent FIR colors similar to M\,81, most of 
the PAH emission should be from the cold component, in agreement
with the results in the current paper. Furthermore, Lu \& Helou (2008)
showed that the fraction of the PAH luminosity that is directly 
correlated with current star formation increases on average as 
FIR color increases.  In contrast, Kennicutt et al. (2009)
showed that nature may present us with a constant fractional
warm PAH or 8\um\ dust emission over a variety of galaxies with 
FIR colors from 0.3 to 1, morphological types of S0 or later, 
and total IR luminosities as high as $10^{11.9}L_{\odot}$.  
With a FIR color of $\sim$0.26, M\,81 is at the cold end of
the color range occupied by the galaxies used in Kennicutt 
\etal (2009). It is therefore of great interest to apply the same 
two-component analysis from this paper to additional galaxies of 
warmer FIR colors.

Many starburst dwarf galaxies show a depressed PAH emission (e.g., 
Engelbracht \etal 2005; Hogg \etal 2005).   The reason could be either 
intrinsic, i.e., a low carbon-based PAH abundance relative to 
silicon-based large dust grains (e.g., Galliano et al.~2008) 
or extrinsic, e.g., via PAH destruction by hard UV or supernova shocks 
associated with massive stars (e.g., Madden \etal 2006; O'Halloran et al.~2006;
Engelbracht \etal 2008).  While the latter explanation 
is more widely accepted today, the former has not been convincingly ruled out. 
It is difficult to differentiate these two scenarios observationally 
because the IR emission in star-forming dwarf galaxies is always 
dominated by their high surface brightness \HII\ regions.
The methods laid out in this paper could in principle be used to 
differentiate these two scenarios:  The mean slope $A$ in eq.~(1) could 
vary from galaxy to galaxy because (i) dust mass spectrum may vary 
(e.g., via metallicity change) or/and (ii) PAH may be destroyed to 
various degrees (as the hardness of UV radiation field may vary).  
In contrast, the mean value of $B$ in eq.~(1) should depend only on 
(i). It is therefore possible to compare statistically samples 
of dwarfs and normal spirals to see if there is a meaningful difference 
in the fitted parameter $B$ between the two galaxy classes.  
Such an analysis 
is now feasible with {\it Herschel} SPIRE data of many low-metallicity
dwarf galaxies (Madden et al.~2013).

\section{Conclusion} \label{sec7}

Using the newly available {\Herschel} images at 250 and 500\um\ and
the existing {\Spitzer} images at 8 and 24\um\ and ground-based 
H$\alpha$ image, we demonstrated a new, two-component correlation
analysis method that is applicable to galaxy images of any angular
resolution and takes into consideration the frequency-integrated 
spectrum of the cold dust emission.  

Using this method, we showed that, in the case of M\,81, the surface brightnesses 
of both the 8\um\ emission from PAHs and the 24\um\ emission from VSGs 
can be quantitatively decomposed into two components, free from any 
model-dependent assumption:  (a) a component that scales with the surface 
brightness of the H$\alpha$ line emission, which is a well known tracer of 
current star formation rate; and (b) a component that scales with 
the surface brightness of the cold, diffuse dust emission at 500\um, 
which is known to be heated primarily by evolved stars.  Roughly, 
the cold components constitute about 67\% and 48\% of the observed 8 
and 24\um\ emissions, respectively.  

We argued that these estimates on the fractional cold dust component likely 
represents a lower limit if any of the following uncorrected systematics in 
our H$\alpha$ image is significant: patchiness in the internal extinction, 
non-stellar origin of the nuclear H$\alpha$ emission, and an elevated 
\NII\ contamination to the H$\alpha$ line flux within the galaxy 
bulge.

\acknowledgements

We thank the anonymous referee for a number of useful comments that 
improved the presentation and clarity of this paper.
This research has made use of the NASA/IPAC Extragalactic Database 
(NED) which is operated by the Jet Propulsion Laboratory, California 
Institute of Technology, under contract with the National 
Aeronautics and Space Administration.
SPIRE has been developed by a consortium of institutes led by
Cardiff University (UK) and including Univ. Lethbridge (Canada);
NAOC (China); CEA, OAMP (France); IFSI, Univ. Padua (Italy); 
IAC (Spain); Stockholm Observatory (Sweden); Imperial College London,
RAL, UCL-MSSL, UKATC, Univ. Sussex (UK); and Caltech/JPL, IPAC,
Univ. Colorado (USA). This development has been supported by
national funding agencies: CSA (Canada); NAOC (China); CEA,
CNES, CNRS (France); ASI (Italy); MCINN (Spain); Stockholm
Observatory (Sweden); STFC (UK); and NASA (USA).


\newpage

\end{document}